**Recent developments in the characterization of superconducting films by microwaves**

M.A. Hein[a], M. Getta[a], S. Kreiskott[a], B. Mönter[a], H. Piel[a], D.E.Oates[b], P.J. Hirst[c], R.G. Humphreys[c], H.N. Lee[d], S.H. Moon[d]

[a] Dept. of Physics, University of Wuppertal, Germany, [b] MIT Lincoln Laboratory, Lexington, MA, U.S.A., [c] QinetiQ, Malvern, U.K., [d] LG Electronics Inc., Seoul, Korea.

Abstract: We describe and analyze selected surface impedance data recently obtained by different groups on cuprate, ruthenate and diboride superconducting films on metallic and dielectric substrates for fundamental studies and microwave applications. The discussion includes a first review of microwave data on $MgB_2$, the weak-link behaviour of RABiTS-type $YBa_2Cu_3O_{7-\delta}$ tapes, and the observation of a strong anomalous power-dependence of the microwave losses in MgO at low temperatures. We demonstrate how microwave measurements can be used to investigate electronic, magnetic, and dielectric dissipation and relaxation in the films and substrates. The impact of such studies reaches from the extraction of microscopic information to the engineering of materials and further on to applications in power systems and communication technology.

Address: Dr. Matthias Hein, Dept. of Physics, Univ. of Wuppertal, D-42097 Wuppertal, Germany. Tel. +49(202)439-2747, Fax +49(202)439-2811

Email: mhein@venus.physik.uni-wuppertal.de

## 1. Microwave measurements in condensed matter physics

### 1.1 Theoretical background

This paper illustrates the relevance of microwave measurements for the investigation of electronic and magnetic transport and ordering in condensed matter. While most work has been done on solids, the described techniques could also be adapted to soft matter or liquids. The presentation is based on a selection of examples and does not attempt to provide a complete description of activities in this field.

The complex-valued surface impedance $Z_s = R_s + iX_s$ at a microwave frequency $\omega = 2\pi f$ is related to the permeability and conductivity of a material, $\mu$ and $\sigma$, in the local limit by

$$Z_s = R_s + iX_s = \sqrt{i\omega \frac{\mu(\omega)}{\sigma(\omega)}} \, , \qquad (1)$$

where $R_s$ and $X_s$ are the surface resistance and reactance [1]. The material parameters $\mu$ and $\sigma$ are complex-valued and frequency dependent, e.g., via electronic, magnetic, or dielectric relaxation. The generalized conductivity $\sigma$ characterizes electrical conduction through transport and displacement currents.

The electrodynamic response of *metallic materials* provides information on the electronic properties within a surface layer of the order of the London penetration depth ~200 nm, and on a variable time scale ~$1/\omega$. Such information is complementary to thermodynamic bulk properties and low-frequency transport data. The high frequencies also help to derive Fermi surface properties, which would require strong magnetic fields at very low temperatures otherwise [2]. We will focus in the following on *superconductors*, for which $Z_s$ bears information on the density of states and quasiparticle excitations above and below the transition temperature $T_c$ [3]. It is also related to the phase purity, grain connectivity and interface effects of technical materials, which cannot readily be studied at similar resolution by other techniques [4,5]. Such information is required for optimizing epitaxial films or biaxially textured tapes in terms of critical current density, low loss, and power handling.

Similarly, studies of microwave dissipation and relaxation in *magnetic* or *dielectric* materials have received great attention recently, e.g., to resolve the coupling of spins and charges in magnetic superconductors like the borocarbides or rutheno-cuprates [6,7,8], or materials displaying colossal magneto-resistance [9], or to search for low-loss ferrite or ferro-electric materials for frequency-agile devices [10,11,12].

## 1.2 Brief history of microwave superconductivity

A historical success of microwave measurements was achieved by H. London, when he studied Sn [13]. His investigations led to the first direct experimental determination of the superconducting energy gap, the development of the two-fluid model, and the analysis of the anomalous skin effect [14,15]. Later on, Pippard exploited the information contained in the microwave measurements of noble metals to clarify the Fermi surface of copper [14], and to develop a comprehensive theory of electronic transport in solids [2].

More recently, the characterization of the hole-doped cuprate superconductors revealed a power-law temperature dependence of the quasiparticle fraction at low temperatures, which helped establish the d-wave symmetry of the order parameter [16]. Part of the ongoing work focuses on the high residual $R_s$-values in $Bi_2Sr_2CaCu_2O_{8+y}$, which are difficult to explain merely by gap nodes. Non-quasiparticle $s$-contributions from, e.g., charge density waves in the inherently inhomogeneous cuprates have been suggested [17], which would modify our understanding of electronic transport in these materials. Other microwave studies aim at resolving the symmetry of the order parameter in the electron-doped cuprates [18]. Comparative measurements on $R_{2-x}Ce_xCuO_{4-\delta}$ (R=Pr,Nd) indicated power-law behaviour of $\Delta l(T)$, in accordance with phase-sensitive experiments pointing to d-wave symmetry [19], but in contradiction to tunneling data [20].

The first microwave measurements on the spin-triplet superconductor $Sr_2RuO_4$ at 10 GHz were reported very recently [21]. This material, though isostructural with $La_{2-x}Sr_xCuO_4$ [22], displays the maximum $T_c \sim 1.5K$ in the undoped state. Reflecting the high purity of this material, the surface impedance revealed strong quasiparticle relaxation, $wt \sim 1$, throughout the superconducting state. Such measurements are, beside their relevance to resolve the pair state in the ruthenates, of great general interest since the involved energy scales for frequency, temperature, relaxation and pairing are all comparable.

$MgB_2$ is another novel superconductor of great interest, due to its high $T_c$ and possible applications [23]. Beside polycrystalline pellets and wires, thin films have become available for extended microwave studies (see [24,25] and Sec. 3.1).

At present at least 50 renowned academic and industrial groups study superconductors at microwave frequencies, to about equal parts for fundamental and applied physics. While the future development of this discipline depends on available funding, exciting results can be expected in the near term, both for broadened understanding of superconductivity and applications optimizing our social and economic resources.

## 2. Microwave measurements for the characterization of thin films

Conceptually, the surface impedance describes the propagation of electromagnetic waves across the interface between vacuum and matter. Eq.(1) implies bulk properties, when the thickness of the material extends over many penetration lengths, $t \gg l$. New aspects arise for $t < l$, due to partial reflection and transmission at the additional film-substrate interface, e.g., in very thin films or near $T_c$. The effective surface impedance follows from an implicit equation [5,26,27]:

$$Z_{eff}(t) = Z_{s\infty} \times \frac{Z_{s\infty} \times \tanh(\boldsymbol{b}t) + Z_{sub}}{Z_{s\infty} + Z_{sub} \times \tanh(\boldsymbol{b}t)}, \tag{2}$$

where $Z_{s\infty}$ is the bulk value, $\boldsymbol{b} = i\boldsymbol{wm}/Z_{s\infty}$ the propagation constant, and $Z_{sub}$ the surface impedance of the substrate.

Eq. (2) can be solved approximately, e.g., for a superconducting film of thickness $t_S$ on a lossless dielectric, leading to $Z_{eff}(t_S) \sim Z_{s\infty} \times \coth(\boldsymbol{k}t_S)$, where $\boldsymbol{k}$ is the inverse skin (penetration) depth above (below) $T_c$. Another interesting case is a normal conducting cap layer of thickness $t_N$ and resistivity $\boldsymbol{r}_N$ on a superconducting film. The propagation of the wave through the cap layer into the superconductor, or vice versa, enhances $R_{s\infty}$ in proportion to $t_N$, namely by $X_{s\infty}^2 \times t_N/\boldsymbol{r}_N$ or $X_{s\infty}^2 \times t_N/\boldsymbol{r}_N \times (\boldsymbol{l}/t_S)^2$, respectively. The explicit $\boldsymbol{l}$-dependence in the latter case has been used to extract absolute $\boldsymbol{l}$-values [28].

Eq. (2) becomes more complicated for multilayers, like a superconducting film on a metallic substrate with dielectric buffer. The effective reactance of the buffer can become comparable with the reactance of the superconductor and the resistance of the metal. Numerical techniques and reference measurements are then required to determine $Z_{s\infty}$.

The principle benefit of $Z_s$ measurements for the characterization of thin films is increased by technological considerations. Thin films are often much more relevant for applications than bulk material, in terms of fabrication, refrigeration and power consumption. Thin films are also preferable for highly anisotropic materials like the cuprates, or for materials prepared by vapor diffusion like Nb$_3$Sn [5] or MgB$_2$ [24]. Should the preparation of single-crystals be complicated by an unfavorable thermodynamic phase diagram, thin films may allow to evaluate intrinsic properties. Alternatively, comparative measurements of poly- and single-crystalline bulk and thin film samples are beneficial to study the impact of defects like grain boundaries, interfaces or impurities [5,29,30].

$Z_{eff}$-measurements also provide information on the electrodynamics of the substrate. For instance, the strongly temperature dependent permittivity of $SrTiO_3$ leads to oscillations of $Z_{eff}(T)$ [26]. Present efforts aim at understanding the different microwave properties of bulk and thin film $SrTiO_3$ for device applications. Another example is the observation of dielectric relaxation in $LaAlO_3$ or MgO, which cause non-monotonic temperature and field dependences of the loss tangent and hence the quality factor of microwave devices [31-33].

## 3. Recent developments of the microwave characterization of superconductors

### 3.1 The surface impedance of $MgB_2$

Many groups have been investigating the microwave properties of $MgB_2$ during various stages of material optimization [24,25,34-38]. Fig. 1 compares our first $R_{eff}(T)$-data of a polycrystalline 400nm-thick $MgB_2$ film on sapphire with three other superconductors at 87 GHz. The lowest $R_s$-values in the entire $T$-range are observed for $Nb_3Sn$, due to its large, almost isotropic energy gap and its low normal-state resistivity. Epitaxial $YBa_2Cu_3O_{7-\delta}$ (YBCO) displays the second lowest losses for $T<T_c$, but the large resistivity causes significant corrections of $R_{eff}(T>T_c)$. The $MgB_2$ film displays a low $R_{eff}=0.49$ $\Omega$ in the normal state, which corresponds to a resistivity of $r\sim20$ $\mu\Omega$cm. The broad transition indicates inhomogeneous film properties, but the residual resistance is promisingly low. This becomes clearer from Fig. 2, which summarizes $R_s(f)$-data of $MgB_2$ at three selected temperatures. The temperature dependence of $R_{eff}$ below about 30 K is consistent with a reduced energy gap $\sim0.8$, in accordance with other RF studies [37,39].

The scatter of the $R_s(f)$-data in Fig. 2 reflects the variability of the quality of bulk samples and films, which is not yet intrinsic. Only the lowest $R_{eff}$-values at 4.2 and 20 K scale almost like $f^a$ with $a\sim2$, as expected for a phase-pure superconductor below the gap frequency. The $R_s$-values are well below those of granular YBCO, and already close to epitaxial YBCO at 77 K [5]. The frequency exponent decreases towards $a \sim 1$ for lower sample quality and higher temperatures, indicating the growing influence of normal or weakly superconducting material. The high value of $R_{eff}(30K)$ at 87 GHz could indicate, besides inhomogeneities, the existence of a small gap value.

## 3.2 The surface impedance of YBCO on RABiTS

Microwave measurements of YBCO on rolling-assisted biaxially-textured substrates (RABiTS) [40] are helpful to optimize the grain boundary coupling and phase purity. Fig. 1 shows $R_{eff}(T)$ of a 400nm-thick YBCO film on a 340nm-thick $CeO_2/YSZ/CeO_2$ trilayer on textured Ni [41]. While the low normal-state resistance is related to the low impedance of the thin buffer and the metallic substrate, the high $R_s$-level in the superconducting state is attributed to low- and large-angle grain boundaries or microcracks [5,42-44]. These effects add not only to the total surface resistance but lead also to a larger thickness correction due to an enhanced penetration length.

Fig. 3 illustrates the influence of weak grain boundary coupling, by relating $R_{eff}$ to the critical current density. The scaling $R_{eff}(77K) \propto J_c^{-1/2}$ for all types of YBCO is reminiscent of Josephson coupling or quasiparticle tunneling at grain boundaries. This result is in contrast with the recent findings for epitaxial films [43], probably indicating the different micro-structure of the two types of samples. At 4.2 K, the losses remain on a much higher level in the RABiTS samples than in epitaxial films. The different scaling at low and high temperatures is emphasized by the similarity of $R_{eff}(77K)$ and the difference $R_{eff}(77K) - R_{eff}(4.2K)$. Our results illustrate that high-frequency measurements are more sensitive to defects than low-frequency techniques, which essentially map critical current densities.

3.3 The nonlinear surface impedance of superconductors on MgO

Understanding the nonlinearities in microwave devices is relevant for applications in communication systems [45]. We have therefore studied microwave losses and intermodulation distortion (IMD) of epitaxial YBCO films on MgO using stripline resonators [33,45]. The total surface resistance contains contributions from the superconductor and the dielectric, $R_{tot}=R_s+G\times\tan\boldsymbol{d}$, where $G$ is a geometry factor. Changes of the reactance are similarly composed of changes of the penetration depth and dielectric permittivity.

Fig. 4 compares the power dependent $R_{tot}$ for YBCO and Nb, which display very similar behaviour. We conclude that $R_{tot}$ is dominated by power dissipation in the dielectric and the superconductor at low and high power, respectively, which is in accordance with the linear and quadratic frequency dependences of $R_{tot}$ measured in the two power ranges. Using $R_s\sim2\ \mu\Omega$, which is typical for YBCO at 2.3 GHz and 1.7 K [5], and $R_{tot}\sim20\ \mu\Omega$, we derive $\tan\boldsymbol{d}\sim3\times10^{-5}$, which is 1-2 orders of magnitude higher than expected for MgO at 20 K [46,47].

Fig. 5 relates $R_{tot}$ of the YBCO sample to $\Delta X_{tot}$ and the IMD signal at 5 K and 2.3 GHz. $R_{tot}$ passes through a minimum at $-20$ dBm, while $X_{tot}$ remains constant, at least up to $-30$ dBm. The increase of $R_{tot}$ and $X_{tot}$ at high power is attributed to the nonlinear response of the superconductor [5]. The IMD signal at low power displays approximately the expected cubic power dependence [48], passes through a plateau in the region where $R_{tot}$ decreases, and rises again more steeply where $R_{tot}$ and $X_{tot}$ increase. The correlation between $R_{tot}$ and the IMD product reveals the dissipative nature and a fast response time of the anomaly.

We were able to model the behaviour by defect dipole relaxation in MgO [33,49]. Our data imply $\boldsymbol{w}\boldsymbol{t}_d<1$ and a electric field-dependence of the dielectric relaxation time $\boldsymbol{t}_d$ on a scale of $E_0\sim100$ V/m at $T<5$ K. This value increased to $\sim30$ kV/m at 20 K, reflecting the growing influence of the superconducting nonlinearity, which eventually masked the dielectric anomaly at elevated temperatures.

## 4. Conclusions

We have illustrated the benefit of surface impedance measurements for the characterization of cuprate, ruthenate and diboride superconducting films on metallic and dielectric substrates for fundamental studies and applications by selected examples. The high frequencies make this technique sensitive to electronic, magnetic, or dielectric dissipation and relaxation in the films and substrates.

The temperature and frequency dependent surface resistance of polycrystalline $MgB_2$, while presently limited by the quality of the available material, is already comparable to that of optimized epitaxial YBCO films, hence opening great potential for applications.

The surface resistance of biaxially textured YBCO films on buffered Ni reveals strong limitations from weak Josephson coupling across grain boundaries or microcracks. Due to the high sensitivity of microwave measurements, $R_s$-guided optimization of superconducting RABiTS tapes is expected to facilitate further development towards conductors with high engineering critical current densities.

The temperature and power dependent surface resistance of superconducting films on MgO displayed pronounced anomalies, which are associated with defect dipole relaxation in the substrate. The nonlinear loss tangent of MgO is relevant for any microwave device application involving this dielectric.

## Acknowledgments

We thank S.Y. Lee, S. Ohshima, and T.C. Shields for providing $R_s$ data on $MgB_2$ and YBCO prior to publication, and J. Derov, J. Halbritter, P. Lahl, N. Newman, S.H. Park, A.

Velichko, and R. Wördenweber for helpful discussion. This work has been funded in part by the EPSRC and MOD (UK), and AFOSR (U.S.A.). Part of this material is based upon work supported by the European Office of Aerospace Research and Development, AFOSR/AFRL, under Contract F61775-01-WE033.

Figure captions

Fig. 1. Temperature dependent surface resistance of epitaxial YBCO on LaAlO$_3$ (black dotted curves), polycrystalline Nb$_3$Sn on sapphire (grey dotted), biaxially textured YBCO on buffered Ni (dashed), and polycrystalline MgB$_2$ on sapphire (solid). The thick (thin) curves denote effective data (corrected for film thickness).

Fig. 2. Frequency dependent $R_{eff}$ of bulk (small black symbols) and thin film MgB$_2$ (large grey) at 4.2 K (solid squares), 20 K (hatched), and 30 K (shaded), collected from [24,25,34-38] and unpublished sources. The grey dotted line represents $R_s$(77K) of epitaxial YBa$_2$Cu$_3$O$_{7-\delta}$ films [5].

Fig. 3. Correlation between $R_{eff}$ of biaxially textured YBCO films on Ni at 4.2 K (solid squares) and 77 K (shaded) and $R_{eff}$(77)−$R_{eff}$(4.2) (hatched) and the critical current density at 77 K. The dotted line denotes $R_{eff} \propto J_c^{-1/2}$. Data on electrophoretic YBCO films on Ag and epitaxial films on LaAlO$_3$ [5] are shown for comparison.

Fig. 4. Power dependent $R_{tot}$ for a YBCO (diamonds) and a Nb film (triangles) on MgO [33]. Absolute $R_{tot}$-values were 8.5 and 12 μΩ at 5 K, and 19 μΩ for both samples at 1.7 K.

Fig. 5. Correlation between $R_{tot}$ of the YBCO film of Fig. 4 (circles), total reactance (triangles) and IMD product (squares) [33].

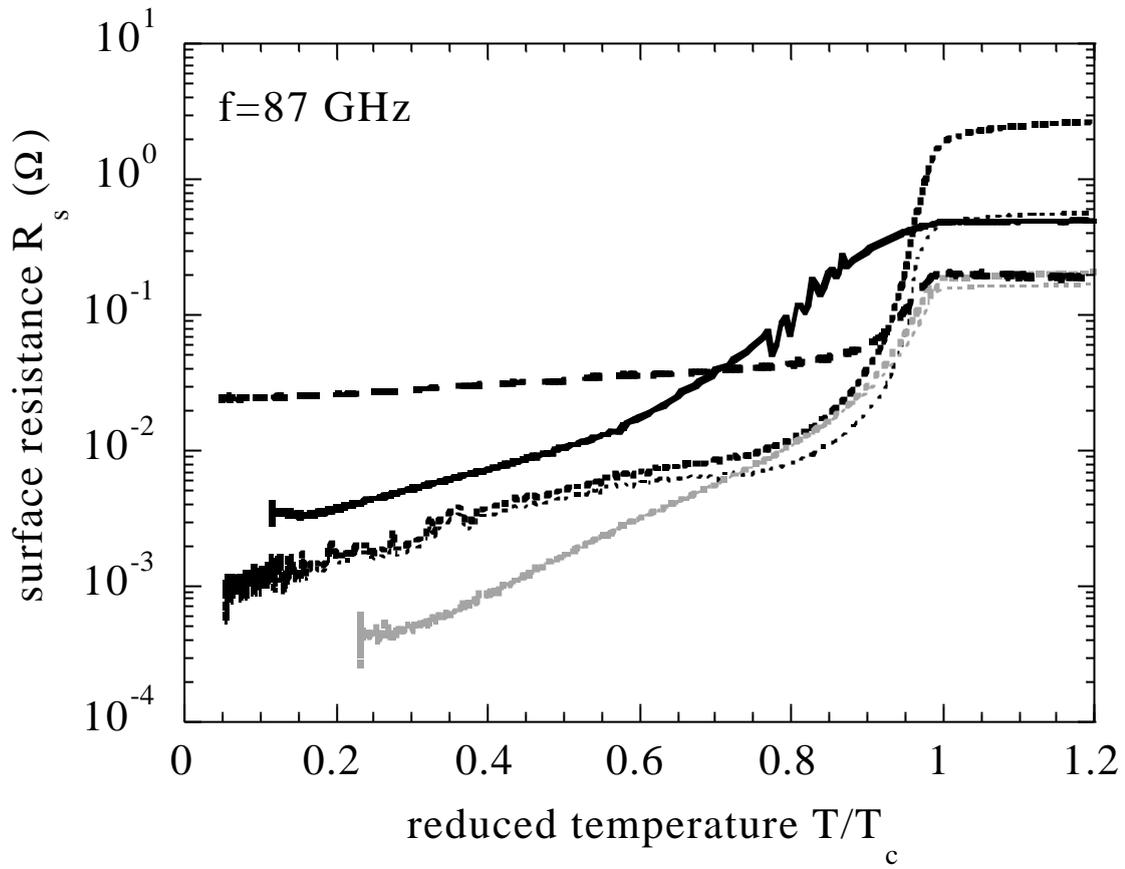

Fig. 1, Hein et al., paper B3-01.

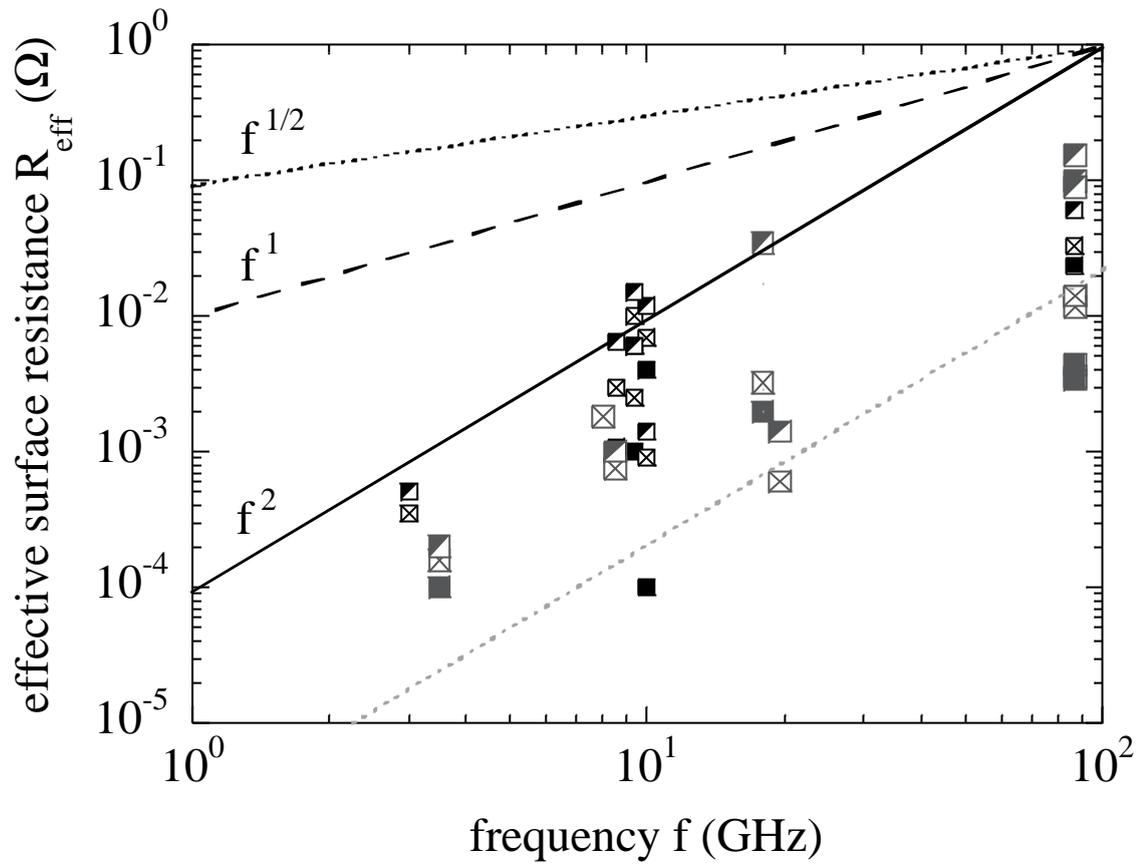

Fig. 2, Hein et al, paper B3-01.

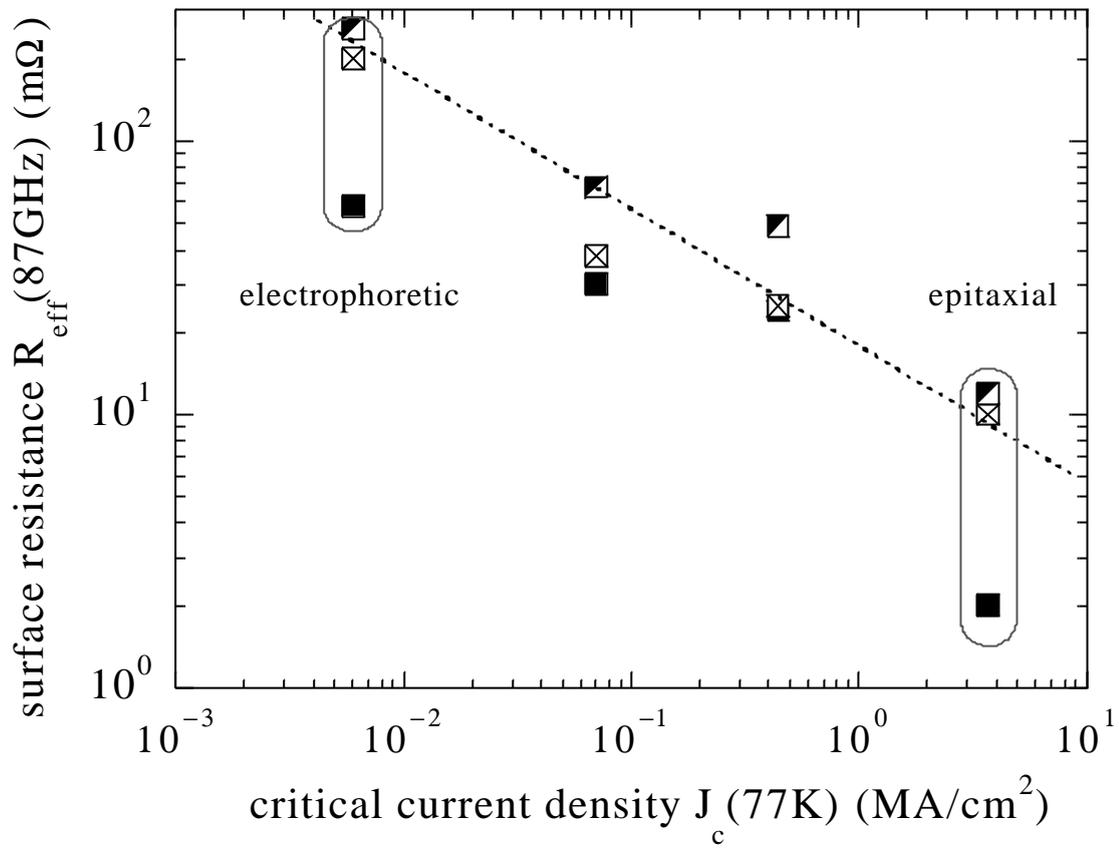

Fig. 3, Hein et al., paper B3-01.

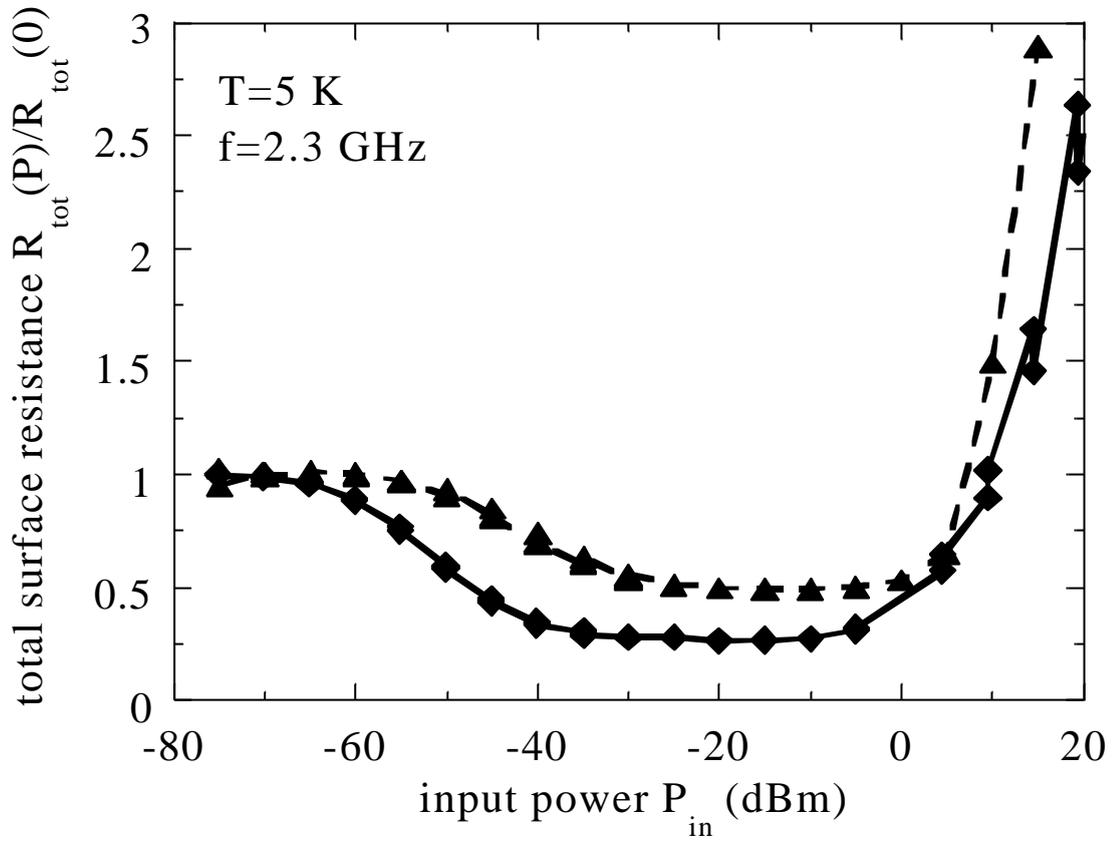

Fig. 4, Hein et al., paper B3-01.

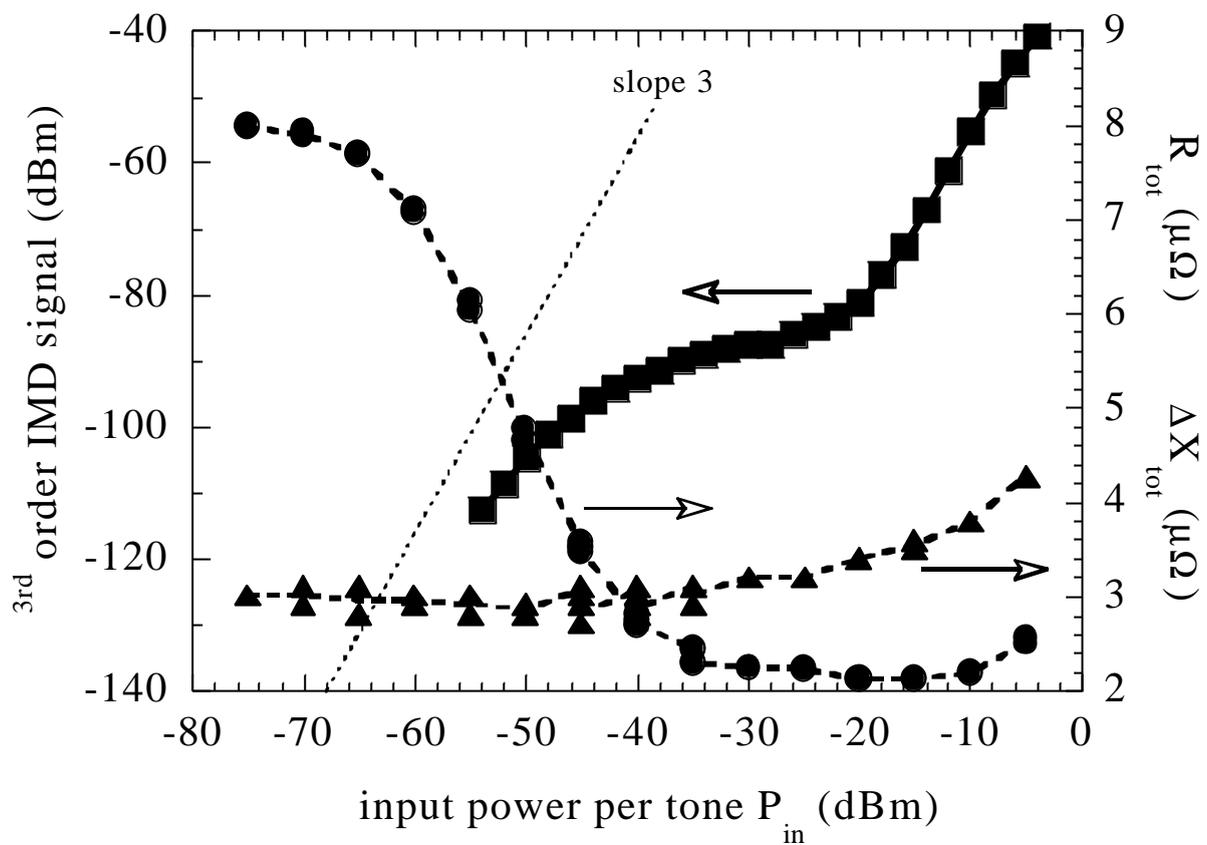

Fig. 5, Hein et al., paper B3-01.